\documentclass[12pt,preprint]{aastex}
\begin{document}

\title{Spectrally Dispersed $K$-Band Interferometric Observations of Herbig
Ae/Be Sources: Inner Disk Temperature Profiles}

\author{J.A.  Eisner\altaffilmark{1,2},  E.I. Chiang\altaffilmark{1,3}, 
B.F. Lane\altaffilmark{4,5}, \& R.L. Akeson\altaffilmark{6}}
\altaffiltext{1}{University of California at Berkeley, 
Department of Astronomy, 601 Campbell Hall, Berkeley, CA 94720}
\altaffiltext{2}{Miller Fellow}
\altaffiltext{3}{Alfred P. Sloan Research Fellow}
\altaffiltext{4}{MIT Kavli Institute for Astrophysics and Space Research,
MIT Department of Physics, 70 Vassar Street, Cambridge, MA 02139} 
\altaffiltext{5}{Pappalardo Fellow}
\altaffiltext{6}{Michelson Science Center, California Institute of Technology,
MC 100-22, Pasadena, CA 91125}
\email{jae@astro.berkeley.edu}


\begin{abstract}
We use spectrally dispersed near-IR interferometry data to constrain the
temperature profiles of sub-AU-sized 
regions of 11 Herbig Ae/Be sources.  We find 
that a single-temperature ring does not reproduce the data well.  
Rather, models incorporating radial temperature gradients are 
preferred.  These gradients may arise in a dusty disk, or may reflect
separate gas and dust components with different temperatures and spatial
distributions.  Comparison of our models with broadband spectral energy 
distributions suggests the latter explanation.
%While we see tantalizing hints that gaseous emission may contribute
%to our measurements, it appears that models incorporating only continuum
%emission are sufficient to explain the present data.
%Disk gas, while undoubtedly present, does not seem to contribute
%significantly to our observations of inner disk emission.
The data support the view that the near-IR emission of 
Herbig Ae/Be sources arises from hot circumstellar dust and gas
in sub-AU-sized disk regions.
%We constrain the inner radii, inner disk temperatures, and power-law 
%exponents of the radial temperature profiles of our dust disk models.  
Intriguingly, our derived temperature gradients appear systematically steeper
for disks around higher mass stars.  It is not clear, however, whether this 
reflects trends in relative dust/gas contributions or gradients within
individual components.
\end{abstract}

\section{INTRODUCTION \label{sec:intro}}
Disks of dust and gas around young stars provide laboratories for studying the 
initial conditions of planet formation.  A knowledge of disk temperature 
profiles at stellocentric radii $R \la 1$ AU 
is crucial for understanding both terrestrial and giant planet formation.  
Planet formation is only possible beyond the 
dust sublimation radius (where temperatures are 
$\la 1500$ K), since dust is an essential building block of terrestrial planets
and giant planet cores.
The disk temperature profile also dictates the 
location of the ``snowline'' \citep[e.g.,][]{LECAR+06}, beyond which 
temperatures are low enough for water ice formation.  The snowline tells us 
where icy material originates, which is important for understanding
how water-rich, Earth-like planets come to be.  Furthermore, because the 
surface density in solids is higher at and beyond the snowline, giant planet 
formation is facilitated there.

%These disks are also an integral part of the star formation process, and
%a knowledge of how disks transfer angular momentum outwards and matter 
%inwards is crucial for understanding the stellar mass function, the 
%angular momenta of young and old stars, and the generation of jets and
%outflows.
%The temperature profiles of protoplanetary disks
%are crucial in constraining planet formation and disk accretion.
%Most observational studies of protoplanetary disks focus on the dust,
%which is the dominant source of continuum opacity.  
%Broadband spectral
%energy distributions \citep[e.g.,][]{BBB88,HILLENBRAND+92} combined with 
%However, open 
%questions remain about the inner disk temperature structure.  For example,
%the inner disk edge seems to puff-up, due to the direct incidence of stellar
%radiation, and it is unclear how the shadowing of this puffed-up inner edge
%might affect the temperature at slightly larger disk radii.

Spectral energy distributions (SEDs) provide some information about disk
temperature profiles \citep[e.g.,][]{HILLENBRAND+92,CG97}, but cannot provide
unambiguous constraints.  The addition of spatially resolved information 
breaks degeneracies inherent in SED modeling, and 
enables mapping of the density and temperature structure of disks
\citep[e.g.,][]{EISNER+05,VANBOEKEL+04,WILNER+00}.  
For example, near-IR interferometric observations have shown that 
simple geometrically thin disk models \citep*[e.g.,][]{LP74,ALS87} 
generally do not fit the data obtained
for protoplanetary disks, while the inclusion of puffed-up inner disk edges 
\citep[e.g.,][]{DDN01} allows models to match the data well for most sources
\citep[e.g.,][]{EISNER+04,EISNER+05}.

To further constrain inner disk temperature profiles, spatially resolved 
observations at multiple wavelengths are necessary. Since dust (and gas) 
emission is temperature and wavelength dependent, different radial disk 
temperature profiles produce distinct spatial distributions of 
emission at different wavelengths.  Thus, measurements of
emission intensity and size as a function of wavelength can be used
to determine inner disk temperature profiles.

Spectrally dispersed interferometry observations also have the potential to 
probe the spatial distribution of spectral line emission relative to  
continuum emission.  Most of the mass in 
protoplanetary disks is contained in gas, which emits primarily through
molecular and atomic transitions.   Investigating spectral line emission
is therefore crucial for understanding disk dynamics and the mass assembly 
and migration of gas giant planets.

Here we present observations that spatially resolve the $K$-band 
($\lambda_0 = 2.2$ $\mu$m; $\Delta \lambda=0.4$ $\mu$m) emission from 
protoplanetary disks around intermediate mass pre-main-sequence stars. These
observations are also spectrally dispersed across 5 spectral channels ($R=25$).
We use these data to investigate the temperature profiles of disks at
stellocentric radii less than 1 AU.  We also explore whether disk
models that include only emission from dust are sufficient to explain our 
observations or whether gaseous emission, including spectral line emission from
hot water vapor or CO, needs to be considered.  
%Because emission lines from hot water vapor and CO 
%occur in only some of the spectral channels, we can use any variation of 
%measured size with wavelength to constrain the relative radii of dust and gas 
%emission.  This task is difficult with only 5 spectral channels, and we 
%therefore consider the results presented here to be preliminary; 
%interferometers with higher spectral resolution will soon facilitate the
%spatial resolution of gaseous emission.

\section{OBSERVATIONS AND CALIBRATION \label{sec:obs}}
The Palomar Testbed Interferometer (PTI) 
is a long-baseline near-IR Michelson interferometer 
located on Palomar Mountain near San Diego, CA \citep{COLAVITA+99}.
PTI combines starlight pairwise from three 40-cm aperture telescopes using
a Michelson beam combiner, and the resulting 
fringe visibilities provide a measure of the brightness distribution
on the sky (via the van Cittert-Zernike theorem).  PTI measures
normalized squared visibilities, $V^2$, which provide unbiased estimates of
the visibility amplitudes \citep{COLAVITA99};  $V^2$ is unity for point 
sources and smaller for resolved sources.

We observed 14 Herbig Ae/Be (HAEBE) 
sources with PTI between May 2002 and January 2004.
Properties of the broadband $K$ emission were reported previously in 
\citet{EISNER+03,EISNER+04}.  Here we consider the subset of 11 objects that 
were spatially resolved (but not over-resolved) in the previous observations, 
and use the spectrally dispersed PTI data to examine trends in the visibilities
with wavelength across a 5-channel ``spectrum.''  
The K-band (2.2 $\mu$m) measurements were dispersed into 5 
spatially filtered spectral channels
on an 85-m North-West (NW)  baseline for all 11 objects,
on an 86-m South-West (SW) baseline for 8,
and on a 110-m North-South (NS) baseline for 5.
The NW baseline is oriented $109^{\circ}$ west of north and has a fringe 
spacing of $\sim 5$ mas, the SW baseline is $211^{\circ}$ west of north
with a $\sim 5$ mas fringe spacing,
and the NS baseline is $160^{\circ}$ west of
north and has a fringe spacing of $\sim 4$ mas.

Visibility data were measured in 130-second ``scans,'' which each consist of
5 equal time blocks.  $V^2$ for the scan is given by the mean of these blocks,
and the measurement uncertainty is the standard deviation.
Measured $V^2$ for the science targets were calibrated by observing sources of 
known angular size and parallax 
(see Eisner et al. 2004 for details of calibrator sources).
These calibrators were unresolved by the interferometer, 
and were close (within $\sim 10^{\circ}$)
to the target sources on the sky;  when calibrating a target 
source $V^2$, data from different calibrators were weighted according to their 
proximity to the target in time and angle \citep[e.g.,][]{BODEN+98}.  

Based on previous observations of binary objects, the uncertainty in 
calibrated, broadband $K$-continuum $V^2$ measurements for PTI is 
$\sim 0.01-0.05$.  The data in the narrow-band spectral channels typically have
similar uncertainties.  We verified this by calculating the standard deviation
of all $V^2$ measurements in each spectral channel
(calibrated as described above) for an unresolved 
``check star'' whose $K$ magnitude is similar to those of our targets.
The standard deviations of the $V^2$ values for the check star in
spectral channels centered at 2.0, 2.1, 2.2, 2.3, and 2.4 $\mu$m are
0.05, 0.03, 0.02, 0.01, and 0.01, respectively.  The larger values at the
shortest wavelengths arise because absorption by
atmospheric water vapor leads to lower fluxes. 
We adopt these values as the noise floors in each
spectral channel for our science data.\footnote{The 
results discussed below do not change substantially
if higher or lower noise floors are used.  As expected, however,
the uncertainties in fitted parameters are larger and the reduced $\chi^2$
values of best-fit models are smaller if a higher noise floor is assumed
(and vice versa for a lower noise floor).}  

We also discard a small amount of data for which low photon counts prevent the 
measurement of any meaningful signal.  For very low photon counts
the measured $V^2$, which are unbiased, may be  $>1$ or $<0$ because of 
large scatter \citep{COLAVITA99}.  The discarded data are all from 
the shortest-wavelength channel, in which
atmospheric water vapor absorption leads to lower photon counts.  
We eliminate all visibilities in the 2.0 $\mu$m channel from nights where the 
majority of the $V^2$ measurements in this channel are $>1$ or $<0$.

\begin{figure}
\plotone{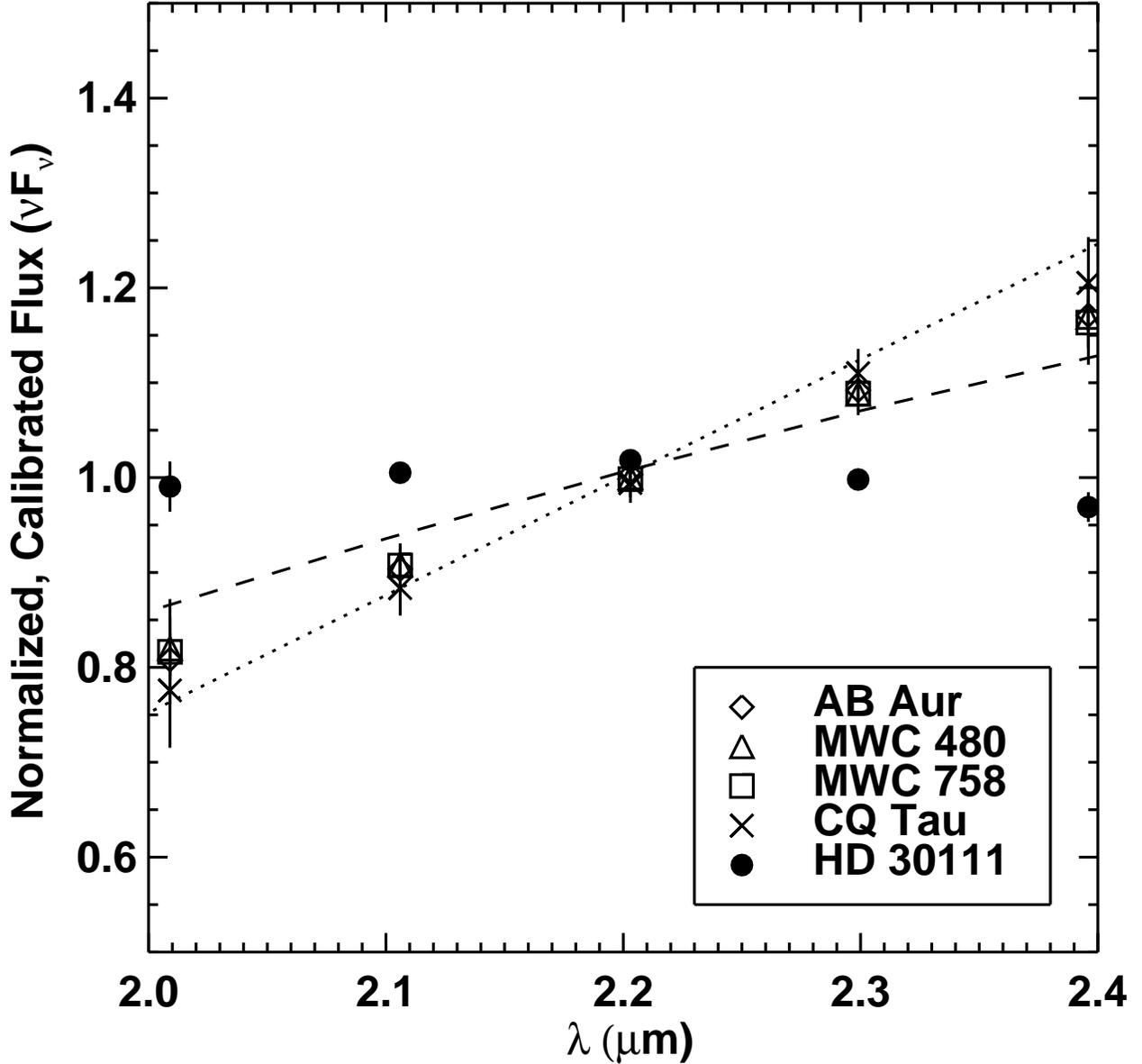}
\caption{Normalized, calibrated photometry for four Herbig Ae/Be stars and
one  ``check star,'' HD 30111.   
We display these objects because the 
photometry for all sources was calibrated using the same set of three
main-sequence calibrator stars, allowing examination of systematic calibration 
effects.  The check star, which has a spectral type of G8III 
($H-K \sim 0.1$ from 2MASS), exhibits
a nearly flat spectrum across the $K$-band, illustrating that our calibration
is reliable.  The photometry for our target sources shows no evidence of 
narrow-band emission features, and the observed spectral slopes for these 
objects are comparable to those of blackbody disk models 
where the temperature of the hottest material is between 1200 K (dotted line) 
and 1600 K (dashed line).
\label{fig:photcheck}}
\end{figure}

We derive crude spectra from 
the observed count rates in our PTI data.  We first normalize the photometry 
by dividing by the mean flux for all spectral channels, and then
calibrate instrumental and atmospheric effects for this normalized photometry 
with the same stars used to calibrate the $V^2$ measurements.
These main-sequence calibrator stars have spectral types ranging from F7 to 
A0, and nearly flat near-IR spectra ($H-K \la 0.2$ based on 2MASS photometry).
Since most calibrators are not colorless A0V stars, the calibrated fluxes for 
the target stars may have some residual 
slopes.  However, this does not appear to be a significant effect. 
Figure \ref{fig:photcheck} shows the fluxes for four of our target stars and 
one  ``check star,'' all calibrated with the same set of 
calibrators.    The check star exhibits a nearly flat spectrum, as expected 
based on the near-IR colors ($H-K \sim 0.1$ from 2MASS photometry).  Thus, our 
differential spectral calibration appears reliable to 
within $\sim 10\%$.  
We estimate (conservatively) that relative (channel-to-channel)
fluxes for our target objects have 
uncertainties of 20\%.  If we remove the continuum slopes of our targets, no 
significant flux variations are observed above the estimated 20\%
uncertainties.

One of our sample objects, MWC 1080, has a nearby companion with 
$0\rlap{.}''78$ separation and $\Delta K=2.70$ \citep{EISNER+04}.  
Since we do not know the 
color of the companion star across the $K$ band, we do not attempt to subtract
contributions of the companion from our flux and $V^2$ data.  Given the
flux ratio of the two stars this may introduce additional uncertainties of 
$\sim 10\%$ in the modeling for this source.  None of our other targets have
known companions which would influence the visibility data at the few percent
level \citep{EISNER+04}.

\section{MODELING AND RESULTS}
Before modeling the circumstellar excess emission from our sample, we
account for the effects of flux from the unresolved central star on 
the measured $V^2$ in \S \ref{sec:ratios}.  In \S \ref{sec:rings},
we fit a simple geometrical model to the $V^2$ data for each spectral channel
individually; our intent is to provide 
a simple picture, with few assumptions, of
how the emission size depends on wavelength.  In \S \ref{sec:mods}, we 
then analyze as a whole our entire dataset of 
spectrally dispersed flux and $V^2$
measurements from 2.0--2.4 $\mu$m
(\S \ref{sec:obs}), fitting disk models that account only for
continuum emission to all of the data simultaneously.  We consider 
a single-temperature ring model, a two-ring model, and a disk model for which 
the temperature decreases smoothly with radius. 
%These models are meant to approximate the following physical configurations,
%respectively: the puffed-up inner rim of a dusty disk; hot gaseous emission
%surrounded by the inner edge of a dusty disk; and a geometrically thin
%irradiated/accretion disk.
In \S \ref{sec:models} we investigate the possible effect of spectral line 
emission on the spectrally dispersed data.  Finally we compute broadband
spectral energy distributions (SEDs) for our best-fit models 
and compare these to observed SEDs from the literature in \S \ref{sec:seds}.

\subsection{Circumstellar-to-Stellar Flux Ratios \label{sec:ratios}}
In order to model the circumstellar emission, we 
must first estimate the contribution to 
the measured fluxes and $V^2$ from the unresolved central stars.  
Following previous investigators \citep[e.g.,][]{MST01}, we use spectral
decomposition to estimate the relative fluxes of the unresolved star and
the circumstellar excess.  We determine the
stellar flux for our sample objects by fitting a blackbody to optical 
photometry (from the literature; see Eisner et al. 2004), and then 
extrapolating to the $K$-band.  The excess flux is given by the difference
between the observed flux and the estimated stellar flux.  To convert the 
normalized 2.0--2.4 $\mu$m fluxes in each channel 
(\S \ref{sec:obs}) into absolute flux units,
we use continuum $K$-band photometric flux measurements from \citet{EISNER+04}.
Derived circumstellar-to-stellar flux ratios for each
spectral channel are listed in Table \ref{tab:fratios}.  Because the
stellar flux declines precipitously with wavelength while
the excess flux increases with wavelength, the circumstellar-to-stellar
flux ratio increases steeply across the $K$-band.

As discussed in \citet{EISNER+04}, uncertainties in the $K$-band photometric
measurements and uncertainties in optical photometry lead to errors in the
estimated circumstellar-to-stellar flux ratios of $\sim 10$--20\%.
In addition, uncertainties in the channel-to-channel flux measurements 
(\S \ref{sec:obs}) introduce errors in the circumstellar-to-stellar flux ratios
of $\sim 10$--20\%.
While many of our sample objects are variable at near-IR wavelengths 
\citep[e.g.,][]{SKRUTSKIE+96}, the photometric K-band measurements 
were nearly contemporaneous with our PTI observations, and thus source
variability should not be a major source of uncertainty.  
Thus the total uncertainties in estimated circumstellar-to-stellar flux ratios
are $\sim 20$--30\%.  These uncertainties lead to small errors in the fitted 
angular sizes because the near-IR emission from our sample is 
dominated by the circumstellar component;
typical uncertainties on fitted parameters are a few percent 
\citep[see][]{EISNER+04}.

\subsection{Emission Radius versus Wavelength \label{sec:rings}}
Before considering the complete dataset, which consists of $V^2$ and fluxes
measured in 5 spectral channels from 2.0 to 2.4 $\mu$m, we provide an
estimate of angular size as measured in each spectral channel.  We employ a 
simple model for the circumstellar excess emission, and fit this
model to the $V^2$ measured in each spectral channel.  Each source was
observed multiple times over multiple baselines, and we fit all of the
$V^2$ measurements in a given channel simultaneously.

Our chosen model is a uniform ring of fixed fractional width. 
%which produces visibilities 
%similar to those for a puffed-up inner disk \citep[e.g.,][]{EISNER+04,DDN01}.
The normalized $V^2$ of a ring, viewed face-on, is given by
\begin{equation}
V^2_{\rm ring} = \left(\frac{2 \lambda}{\pi \theta_{\rm in} r_{\rm uv}
(2f + f^2)} \left\{
(1+f) J_1\left(\frac{[1+f] \pi \theta_{\rm in} r_{\rm uv}}{\lambda}\right) - 
J_1\left(\frac{\pi \theta_{\rm in} r_{\rm uv}}{\lambda}\right) 
\right\} \right)^2,
\label{eq:ring}
\end{equation}
where $\theta_{\rm in}/2$ is the inner angular radius of the ring, 
$1+f$ is the ratio of the outer ring radius to the inner ring radius,
$r_{\rm uv} = (u^2 + v^2)^{1/2}$ is the ``uv radius,'' $u,v$ are the
projected baseline lengths on the sky, and $J_1$ is a
first-order Bessel function.   We take $f=0.2$ for all 
objects.
%,\footnote{This width is compatible with expected rim widths for 
%puffed-up inner disk models \citep{IN05}.}
The only free parameter in the model is $\theta_{\rm in}$.

Inclined geometries are included in this model, and in those discussed below, 
using a transformation to a circularly symmetric coordinate system
\citep[see e.g.,][]{MST01,EISNER+03}.  We assume
values for the inclination and position angle based on previous modeling of
the broadband $K$ emission \citep{EISNER+04}.
%\footnote{For VV Ser,
%we adopt the values of inclination and position angle from \citet{EISNER+03},
%rather than those from \citet{EISNER+04}. As discussed in \citet{EISNER+04},
%there are several reasons why we prefer the values from the earlier work.}  
We verified that these values
were reasonable for the spectrally dispersed data by fitting ring models where
inclination and position angle (in addition to $\theta_{\rm in}$)
were free parameters.  For all objects, the best-fit position angles and
inclinations were consistent within the 1$\sigma$ uncertainties 
across the 5 spectral channels.  Furthermore, these values were compatible
with those derived previously from wide-band $V^2$ 
measurements \citep{EISNER+04}.
We therefore fix the inclinations and position angles in the modeling
presented here.  

\begin{figure}
\plotone{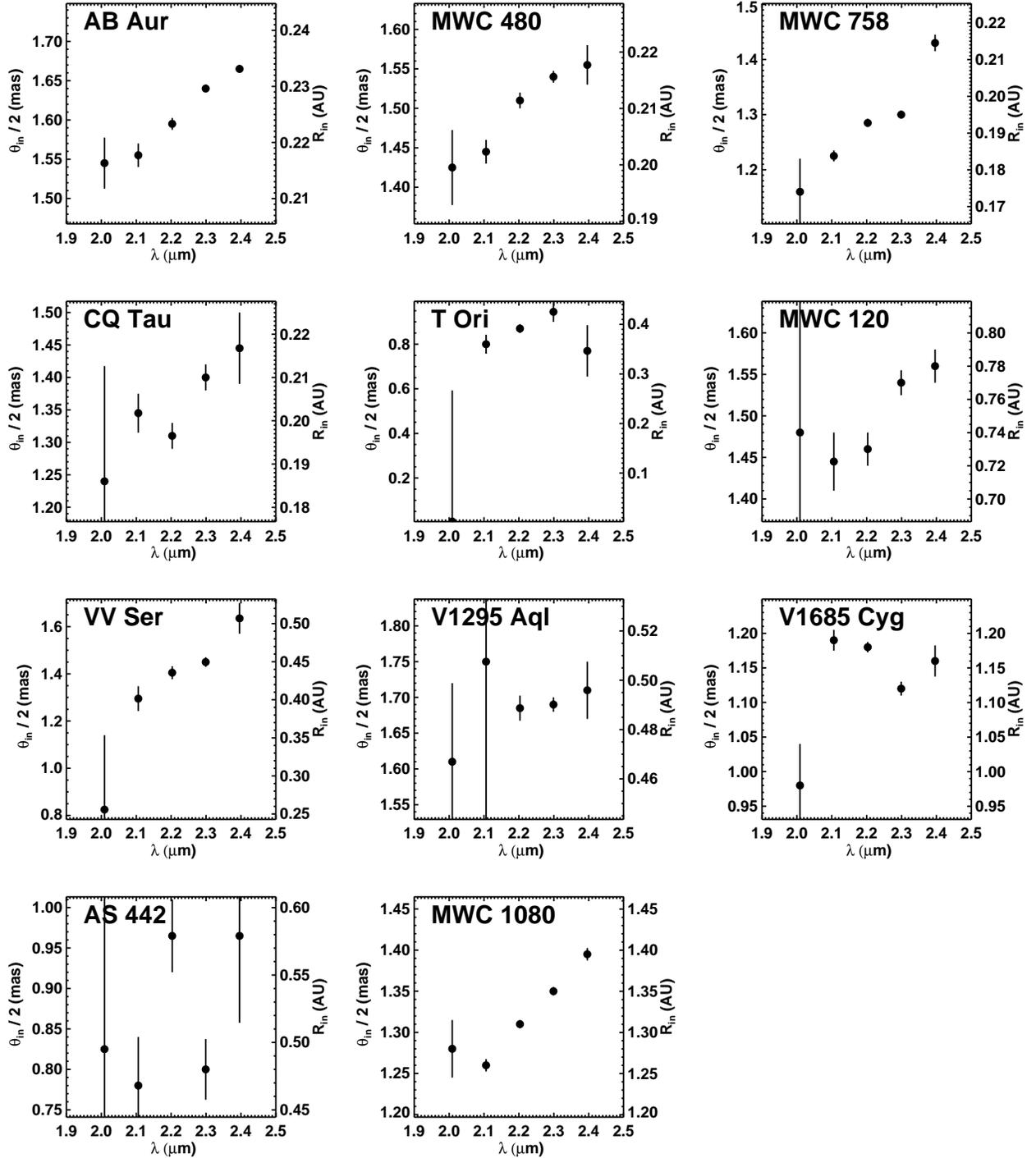}
\caption{Inner ring radii, $\theta_{\rm in}/2$, as a function of wavelength, 
as determined from inclined uniform ring models fitted 
to the data in individual 
spectral channels.  We also provide the linear inner ring radii, $R_{\rm in}$,
computed using assumed distances for these objects 
\citep[][and references therein]{EISNER+04}.  Inclinations, position angles, 
and ring widths were assumed (\S \ref{sec:rings}), and
the inner ring radius was the only free parameter in the model fitting.
%The uncertainties plotted here have not
%been scaled by $\sqrt{\chi_r^2}$, and thus small uncertainties do not
%necessarily indicate a good fit.
\label{fig:sizes}}
\end{figure}

Fitted inner ring radii as functions of wavelength are shown in 
Figure \ref{fig:sizes}.  We also indicate the corresponding
linear radii computed using the distances assumed in 
\citet{EISNER+04}.  Clearly, size increases with wavelength for most 
objects. Such behavior is expected for irradiated disks and/or accretion disks,
where the temperature decreases with increasing stellocentric radius.  
It is not consistent with a single-temperature ring at fixed radius, since 
the angular size of such a ring is independent of wavelength.  
For several objects, the inferred sizes are
not obviously increasing with wavelength.  Rather, the size of the 
circumstellar emission is independent of $\lambda$  within uncertainties
(e.g., V1295 Aql), or is not a monotonic function (e.g., V1685 Cyg).  

\subsection{Circumstellar Emission Models \label{sec:mods}}
To quantitatively investigate the radial temperature structure suggested by 
Figure \ref{fig:sizes}, in this section we fit simple models to our $V^2$ and 
photometry data for all spectral channels simultaneously.  Here we assume
that the models emit only continuum emission; 
in \S \ref{sec:models} we investigate
whether better fits are obtained when spectral line emission is included.
For all models we fit only our $V^2$ and flux data measured from 
2.0--2.4 $\mu$m; in \S \ref{sec:seds} we compute the broadband SEDs predicted
by our best-fit models and compare to observations.

\subsubsection{Single-Temperature Ring \label{sec:modring}}
We first consider the uniform ring model discussed above, but now fit the
model to the complete dataset all at once (instead of fitting each channel
individually).  In addition, we now consider the measured fluxes as well as
$V^2$.  This necessitates the introduction of a ring temperature,
$T_{\rm ring}$, in order to fit the photometry data.
If we assume for simplicity that the ring emits as a blackbody, the flux 
emitted by the ring is
\begin{equation}
F_{\nu}(R)  = \frac{2\pi \cos i }{d^2} B_{\nu} \left(T_{\rm ring} \right) \:  
R_{\rm in}^2 \: \left(f+f^2/2\right) 
\approx \frac{2\pi \cos i}{d^2} B_{\nu} \left(T_{\rm ring} \right) \:  
R_{\rm in}^2 \:f  .
\label{eq:flux-ring}
\end{equation}
Here $B_\nu$ is the Planck function, $d$ is the distance, and $i$ is
the inclination of the ring 
\citep[$d$ and $i$ are assigned from][]{EISNER+04}.  We determine the
values of $R_{\rm in}$ and $T_{\rm ring}$ that provide the best fit to all of
the measured $V^2$ and fluxes for a given source
(Table \ref{tab:rings}). 

\begin{figure}
\plotone{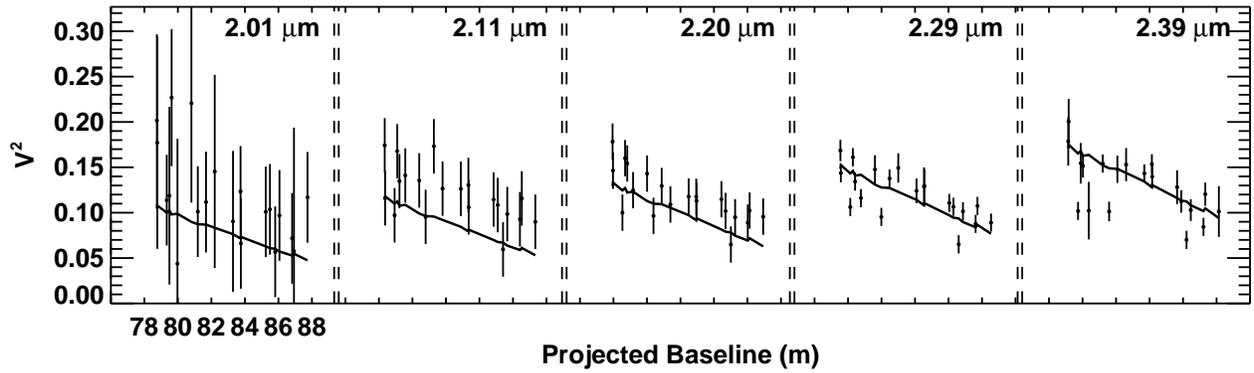}
\caption{PTI $V^2$ data for AB Aur plotted with the best-fit 
single-temperature ring model (\S \ref{sec:modring}).  
The model systematically 
under-predicts the data in the short-wavelength channels.
The reduced $\chi^2$ of the fit for this object, and all other objects in
our sample, is $>1$, indicating poor agreement between this simple model
and the data.
\label{fig:abaur-ring}}
\end{figure}

It is not surprising that the reduced $\chi^2$ values of our best-fit ring 
models are $>1$ (Table \ref{tab:rings}), since the ring model produces the 
same size in each channel, contrary to what is observed in the data
(see \S \ref{sec:rings}).  We illustrate the poorness of fit 
with an example in Figure \ref{fig:abaur-ring}, which shows
the measured $V^2$ along with the prediction of the best-fit 
single-temperature ring model for AB Aur.  While
$\chi_r^2$ is not far from unity for several objects,
this is probably because of large measurement errors 
and does not necessarily imply that the single-temperature ring model 
provides a good physical description of the data.

\subsubsection{Two Rings \label{sec:mod2rings}}
The dependence of source angular size on wavelength suggests that we should 
include temperature-dependent radial gradients in the models.  
While a single-temperature ring
model possesses no such gradients, the addition of a second ring at different
stellocentric radius and temperature will lead to observed source structure 
that is a function of wavelength.  Specifically, the cooler, outer ring
will contribute relatively more emission at longer wavelengths, and thus
the angular size of the two-ring model will tend to increase with wavelength.

We assume the rings have fractional widths of 0.2, and we fix the 
radius of the inner ring at 0.1 AU.  So long as the inner ring is essentially
unresolved by our observations, the exact choice of inner ring radius will not 
substantially alter the quality of the fits.  We verified this by performing
fits where the inner ring radius was left as a free parameter; reduced
$\chi^2$ values did not improve with this extra degree of freedom.
Smaller assumed values of the inner ring radius will, however,
yield higher inner ring temperatures.  Conversely, larger values of the 
inner ring radius give lower inner ring temperatures. However
substantially larger inner ring radii can be 
ruled out since they would be spatially resolved, and would thus lower
the quality of the fits. 

The free parameters of the model are the temperature of the inner ring,
and the radius and temperature of the outer ring. This
is obviously a simple description of a potentially complex system, but will
demonstrate whether a multi-component continuum model is suitable for
explaining our data.  As above, the data fitted include $V^2$ and fluxes
measured from 2.0--2.4 $\mu$m (\S \ref{sec:obs}).

\begin{figure}
\plotone{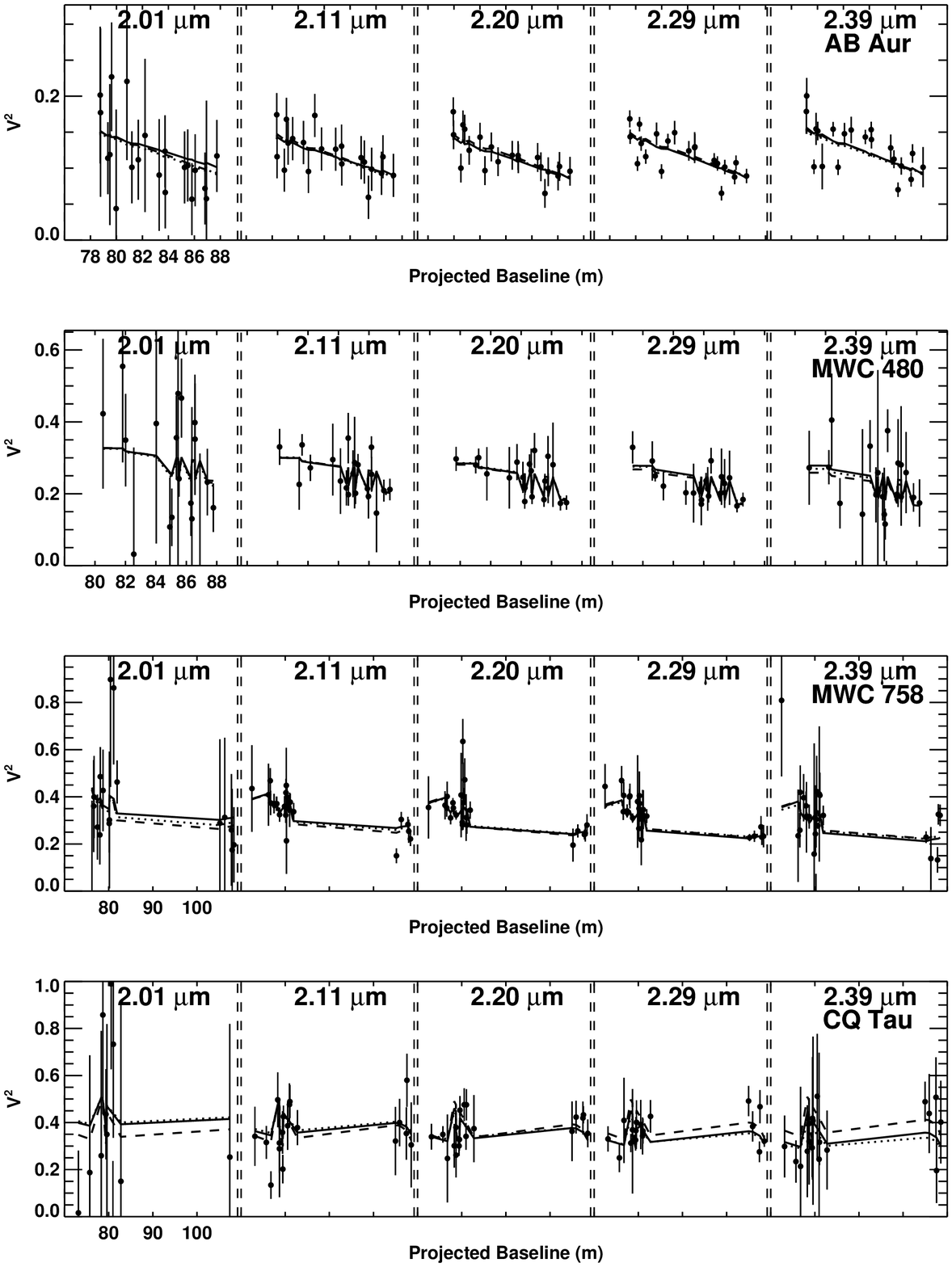}
\caption{PTI $V^2$ data (points), and best-fit models.  Two-ring models
are represented by solid lines. Disk models incorporating smooth
temperature gradients are indicated by dotted lines.  Dashed lines show the 
best fits obtained when only the second and third channels (``continuum''-only 
fits) were used to constrain the fits of these disk temperature-gradient 
models. Data for multiple
baselines, with multiple position angles on the sky (\S \ref{sec:obs}), are
plotted here.  Curves for best-fit models may therefore have a jagged 
appearance since inclined disks have different visibilities when observed
at different position angles.
\label{fig:modfits}}
\end{figure}

$V^2$ values predicted by best-fit two-ring models are shown
in Figure \ref{fig:modfits}. 
Best-fit parameters and reduced $\chi^2$ values are
listed in Table \ref{tab:tworings}.  For objects that display a monotonic 
increase  in size with wavelength (see Figure \ref{fig:sizes}) the reduced 
$\chi^2$ values for this model are lower than for the single ring model 
(\S \ref{sec:modring}).  In contrast, for T Ori, V1295 Aql, V1685 Cyg, and
AS 442, sources for which the size does not appear to increase with wavelength,
the single-temperature ring model provides a lower reduced $\chi^2$.
The high temperature of the inner ring for MWC 1080 is due to
our assumed inner ring radius of 0.1 AU.  A larger inner ring radius---which 
would lead to a lower fitted inner ring temperature---is compatible
with our data;  at the source distance \citep[$\sim 1$ kpc; see][]{EISNER+03}, 
an inner ring radius as large as $\sim 0.5$ AU  would still produce unresolved 
emission as assumed in our model.

\subsubsection{Disk with Smooth Radial Temperature Gradient 
\label{sec:moddisk}}
As a generalization of the two-ring model (\S \ref{sec:mod2rings}), we
construct a disk model with a continuous temperature gradient.
If the temperature decreases over a range of disk radii
then the cooler material at larger radii may contribute relatively 
more emission at longer wavelengths.  We consider a disk model with a
power-law temperature profile,
\begin{equation}
T = T_{\rm in} \left(\frac{R}{R_{\rm in}}\right)^{-\alpha}.
\end{equation}
Here, $R$ is the stellocentric radius, $R_{\rm in}$ is the inner radius
of the disk, and $T_{\rm in}$ is the temperature at $R=R_{\rm in}$.  
The model is described by three free parameters:
$R_{\rm in}$, $T_{\rm in}$, and $\alpha$.  

We assume the model extends from $R_{\rm in}$ to an outer radius
$R_{\rm out}=10$ AU.  The exact choice of $R_{\rm out}$ is not crucial
since most of the 2 $\mu$m emission arises from sub-AU radii.  However
if $R_{\rm out}$ is $\ll 1$ AU  then the model will not produce
substantially different angular sizes at different radii; rather the emission 
would appear to arise from a narrow ring at all wavelengths and the model would
resemble the single-temperature ring model described in \S \ref{sec:modring}.
To explain the dependence of angular size on wavelength, we require 
$R_{\rm out} \ga 1$ AU.

We divide the disk up into annuli and compute the flux and
visibilities for each annulus.
We assume each annulus emits as a blackbody so that the flux in an annulus of 
infinitesimal width $dR$ (observed at the Earth) is
\begin{equation}
dF_{\nu}(R)  = \frac{2\pi \cos i}{d^2} B_{\nu} \left(T \right)  R \: dR ,
\label{eq:fr}
\end{equation}
where $B_\nu$ is the Planck function, $d$ is the distance, and $i$ is
the disk inclination.
The normalized visibilities for an annulus extending from $R_1$
to $R_2 = R_1+dR$ are given by the difference
of visibilities for uniform disks having radii equal to
$R_1$ and $R_2$:
\begin{equation}
V(R) = \frac{\lambda d}{\pi r_{\rm uv} (R_2^2 - R_1^2)}
\left[R_2J_1 \left(\frac{2\pi r_{\rm uv} R_2}{\lambda d}\right) - 
R_1J_1 \left(\frac{2\pi r_{\rm uv} R_1}{\lambda d}\right)\right].
\end{equation}
The sum of $dF_{\nu}$ over all annuli gives the total disk flux, $F_{\nu}$. 
The square of the flux-weighted sum of  $V(R)$ over all annuli gives $V^2$.

The $V^2$ values predicted by
best-fit models including a temperature gradient in the inner disk are shown
in Figure \ref{fig:modfits}.  Table \ref{tab:results} shows that the reduced 
$\chi^2$ values of the best fits are generally $\la 1$.  For
all objects except V1295 Aql and V1685 Cyg
this model provides a better fit (i.e., a lower
reduced $\chi^2$ value) than the single-temperature ring model.
Moreover, the 
temperature gradient model yields a better fit to the data than the 
two-ring model for all sources.  Thus, our spectrally dispersed interferometry
data favor disk models that incorporate radial temperature gradients.
How well the temperature gradient model fits broadband SEDs compared to other
models is discussed in \S \ref{sec:seds}.

For most sources the inner disk sizes and temperatures derived here are 
similar to those inferred by \citet{EISNER+04} from broadband 
(2.0--2.4 $\mu$m) $V^2$ measurements. This is expected since the broadband
$V^2$ measurements from \citet{EISNER+04} have similar values to the
spectrally dispersed measurements presented here. 
However, for objects with small 
best-fit $\alpha$ values (e.g., MWC 758 or CQ Tau), 
our determined inner radii are smaller than the previous estimates by 
$\la 40\%$.   This discrepancy is due to the fact that \citet{EISNER+04}
assumed $\alpha=0.75$ for their disk models; for the smaller $\alpha$ values
inferred here, the inner edge of the disk must extend further inward
to match the data.  The values of $R_{\rm in}$ determined here are more
reliable than previous estimates since $\alpha$ and $R_{\rm in}$ 
have been determined simultaneously.

%Best-fit values of $\alpha$ are typically $\la 0.5$, smaller than expected for
%geometrically thin accretion disks \citep{LP74}.  While a flared disk produces
%$\alpha = 0.5$ \citep[e.g.,][]{CG97}, flaring is not expected to be important
%at the small radii probed by our interferometric observations. These
%small values of $\alpha$ may indicate that we are actually observing 
%temperature gradients across the puffed-up inner disk edges.  In fact, 
%measurements of size versus wavelength may provide the only way to truly
%measure such temperature gradients.

\subsection{Spectral Line Emission \label{sec:models}}

While disk models including only continuum emission can provide reasonable fits
to our spectrally dispersed interferometry data, several objects---T Ori,
V1295 Aql, V1685 Cyg, and AS 442---do not show the monotonic increase of size 
with wavelength expected for the continuum models with temperature gradients 
(\S \S \ref{sec:mod2rings}--\ref{sec:moddisk}).  This is particularly striking
for V1685 Cyg, where the smaller uncertainties strongly imply
non-monotonic dependence of size on wavelength (see Figure \ref{fig:sizes}). 
One potential explanation is that gaseous emission, with a different spatial 
scale than the dust continuum, contributes to the visibilities in certain 
spectral channels.

%To model the emission in each spectral channel, we must specify the
%wavelengths where gaseous emission contributes.
%We must further specify the gas-to-dust emission ratio in each spectral
%channel; this ratio depends on the strength of the gaseous emission features
%and the spectral filling factor of the emission within a channel.  
%A realistic thermo-chemical model is beyond the scope of this work, and
%we therefore make several simplistic assumptions about the gaseous emission.

\begin{figure}
\plotone{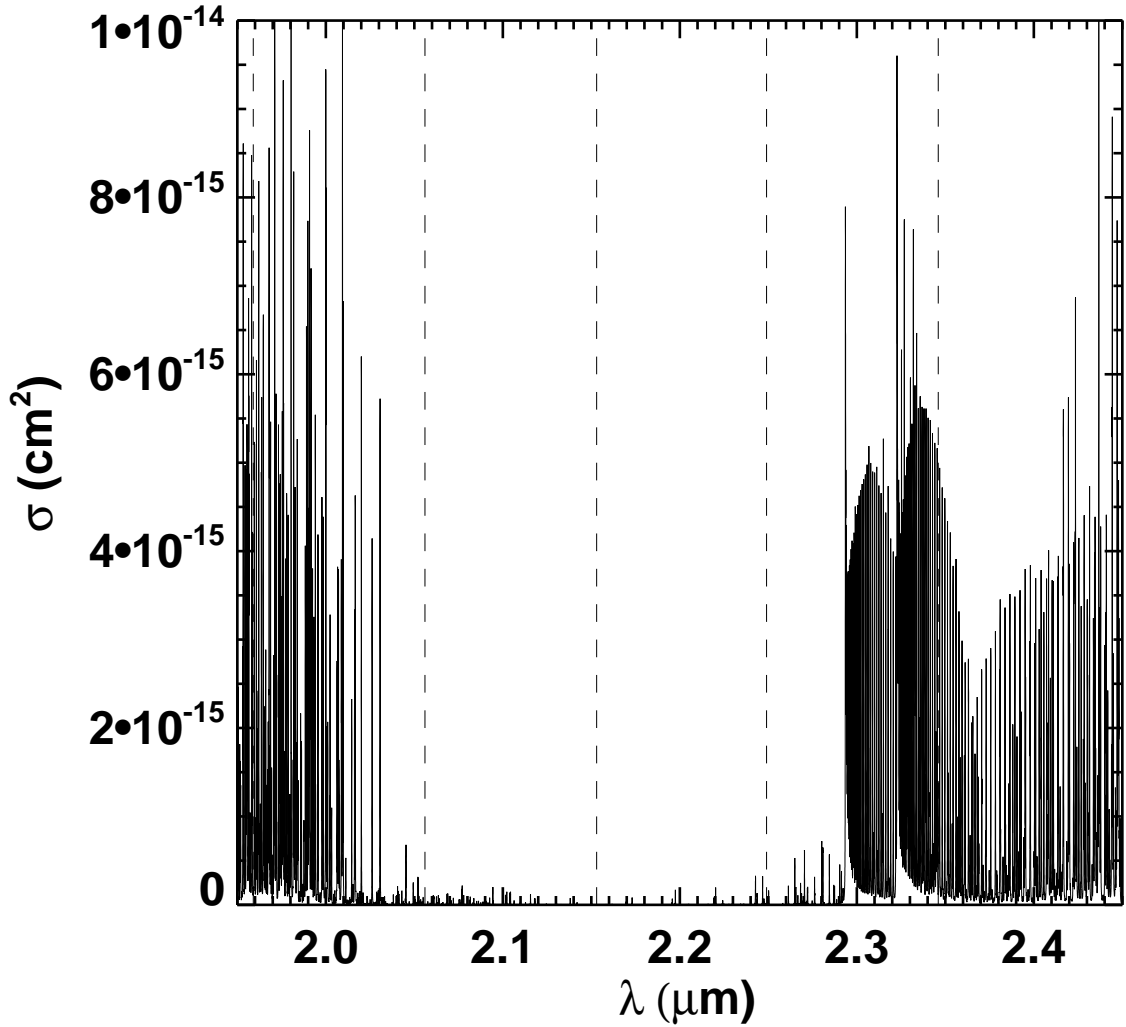}
\caption{Radiative cross sections ($\sigma$) for water and CO
in the $K$-band spectral region (solid line).  Lines are drawn from the 
HITRAN database for temperature $T=3000$ K \citep{ROTHMAN+05}, and 
convolved with Voigt profiles assuming molecular number densities of
$n=10^{19}$ cm$^{-3}$.  
%The product of gas column density and $\sigma$
%yields the optical depth, $\tau$.
The boundaries of the spectral channels used in this 
paper are indicated with dashed lines.
\label{fig:hitran}}
\end{figure}

Based on compilations of H$_2$O and CO emission lines from the
HITRAN database \citep{ROTHMAN+05}, gaseous spectral line emission, if present,
contributes principally to the PTI measurements in the channels centered at 
2.0, 2.3, and 2.4 $\mu$m (Figure \ref{fig:hitran}). 
Previous observations of hot gas near a young star illustrate that gaseous
emission features may be quite broad \citep{CTN04}, and thus this emission
might fill large parts of some spectral channels.

If the dust and gas emission components have different spatial distributions, 
then the visibilities in spectral channels where gaseous emission is important 
(2.0. 2.3, and 2.4 $\mu$m) may be offset from the visibilities in spectral
channels dominated by dust continuum emission (2.1 and 2.2 $\mu$m).
This effect has been observed previously in evolved late-type stars, 
where molecular layers above the stellar photosphere lead to different measured
sizes across the $K$-band \citep[e.g.,][]{TCV02}.
To investigate whether gaseous emission is contributing to our measured 
visibilities, we fit the models discussed above
to only the visibilities measured from 2.1 to 2.2 $\mu$m, and compare these 
fits to those using the data from all channels.

Best-fit parameters for ``continuum''-only fits (i.e., fits using only the 
data measured from 2.1 to 2.2 $\mu$m) are included in Table \ref{tab:results},
and the predicted $V^2$ values are plotted in
Figure \ref{fig:modfits}.  For all sources, the best-fit inner disk sizes
for ``continuum'' channels only and for all channels are equal within
the 1$\sigma$ uncertainties.  While the reduced $\chi^2$ values of the 
``continuum''-only fits are smaller for AB Aur, T Ori, 
VV Ser, AS 442, and MWC 1080 (Table \ref{tab:results}), 
the differences are small for all sources except AB Aur. 

%The reduced $\chi^2$ values for VV Ser and MWC 1080 are only slightly lower,
%suggesting that gaseous emission may contribute.  Figure \ref{fig:modfits}
%shows only a slight difference between the ``continuum'' fits and the fits
%where all data is used for these objects.  The sizes from 
%``continuum''-only fits are larger than the fits where all data are used;
%if one attributes the difference to gaseous emission, then the emitting gas 
%would be found at smaller stellocentric radii than the dust.  However, given
%that the best-fit sizes are consistent at the 1$\sigma$ level, it is
%and thus it is not possible to make any strong claims for the presence of gas.
The incompatibility of a simple disk model with the observed
size versus wavelength for V1685 Cyg was our main impetus for investigating
gaseous emission. However, even if we consider only channels that are not 
expected to be contaminated by gaseous emission, the data are 
still not fitted well by our disk models (note the large $\chi_{\rm r}^2$ 
values in Table \ref{tab:results}); the circumstellar emission for this 
object may not be represented well by a pure disk model.

The spectrally dispersed interferometry data presented here do not
argue strongly for the presence of gaseous spectral line emission in the inner 
disks around our sample HAEBE stars.  We also do not observe any strong 
gaseous emission features in our crude spectra 
(e.g., Figure \ref{fig:photcheck}).  However the
large uncertainties do not rule out gaseous spectral line emission with 
line-to-continuum ratios $\la 20$\%.  Furthermore, the smaller 
$\chi_{\rm r}^2$ values of continuum-only fits for several objects 
(especially AB Aur) are intriguing.  

\subsection{Spectral Energy Distributions \label{sec:seds}}

We fitted the models described above to measured $V^2$ and fluxes measured
from 2.0--2.4 $\mu$m.  We now compare the broadband fluxes predicted by our
best-fit models to observed SEDs.  This 
comparison is motivated by previous analyses of SEDs and $K$-band 
interferometry \citep[e.g.,][]{MST01,EISNER+04}, which showed that 
single-temperature ring models generally
provided better fits to the data than disks with temperature gradients.

Figure \ref{fig:seds} shows the predicted and observed SEDs for our sample.
Disk models incorporating smooth temperature gradients do not fit the SED data
as well as the two other models for almost all objects;
the SEDs are fitted better by single-temperature ring models or by two-ring 
models.  One exception is MWC 1080, for which a disk model fits the SED data
well; for this source the two-ring model does not fit the SED well 
(Figure \ref{fig:seds}).  
This is consistent with previous arguments that early-type stars like
MWC 1080 are surrounded by geometrically thin disks with 
temperature gradients \citep[e.g.,][]{EISNER+04}.

\section{DISCUSSION \label{sec:discussion}}

The size of the near-IR emission for most of our objects varies as a function 
of wavelength, indicating that a simple, single-temperature
ring model for the emission
is untenable; such a model predicts the same size regardless of observing
wavelength.  We showed above that the $V^2$ and flux data measured from
2.0--2.4 $\mu$m are fitted better by models 
incorporating radial temperature gradients.  
%While it is not surprising that
%single-temperature ring models do not provide an adequate description of real 
%astrophysical systems, we note that this model is often adopted when analyzing
%near-IR interferometric data and has been shown to be consistent with 
%broadband $K$-band visibility measurements \citep[e.g.,][]{MM02,EISNER+03}.  

Early puffed-up inner disk models \citep[e.g.,][]{DDN01} generally
assumed that the inner wall emits as a single-temperature blackbody; this 
model resembled a single-temperature ring.  
Even if temperature gradients within the puffed-up inner rim are included
\citep[e.g.,][]{IN05}, the narrow range of radii over which the emission arises
means that these models still resemble single-temperature rings.  Our data
therefore suggest that puffed-up inner rims alone do not fully represent 
the circumstellar emission from HAEBE sources.

Temperature gradients over a range of disk radii
must be included in the model to fit the observational data presented here.  
The best fits to our spectrally-dispersed interferometry are obtained for 
two-ring models or models with power-law disk temperature profiles.
Comparison of our models with broadband SEDs suggests that two-ring models
are preferable.  We suggest that our two-ring model represents a puffed-up edge
of the dust disk and hotter inner disk gas.  The best-fit temperatures of the 
inner ring are substantially hotter than the dust sublimation temperature, 
supporting our hypothesis that
such emission probably traces gas.  The outer rings have 
temperatures less than or comparable to the dust sublimation temperature, 
suggesting that the outer ring emission can arise from dust.

If inner disk/ring temperatures larger than dust sublimation temperatures 
reflect contributions from hot gas emission to measured fluxes and $V^2$, 
we might also expect to detect variations in the visibilities
as a function of wavelength due to gaseous CO or H$_2$O emission features.
In general, no such trends are seen in our data (\S \ref{sec:models}).  
It is possible that the gaseous emission features are so broad that
they produce a ``pseudo-continuum''; there is some support for this suggestion
from previous observations of broad gaseous emission features from a young
star \citep{CTN04}.  It is interesting to note, however, that for AB Aur, 
AS 442, and MWC 1080, objects with 
higher inner disk temperatures, there is a suggestion that the inclusion of
gaseous emission features in the models would improve the fits to the 
observations, since the ``continuum''-only models yield
lower reduced $\chi^2$ values than the fits where all the data is used
(\S \ref{sec:models}).
Future observations with better spectral resolution are 
needed to probe potential spectral line emission in more detail.

While a two-ring model provides the best match to the combination of 
spectrally dispersed and  SED data, a disk model with a radial temperature 
gradient provides the best fit to the spectrally dispersed interferometry data
(from 2.0--2.4 $\mu$m)
for all sources.  Our results suggest that a multi-component model that 
includes temperature gradients in one or more components may provide better
fits to the data than the simple models considered here.  We note that 
MWC 1080, the most massive star in our sample, provides an exception:
disk models with smooth temperature gradients fit all of the data well.

\begin{figure}
\plotone{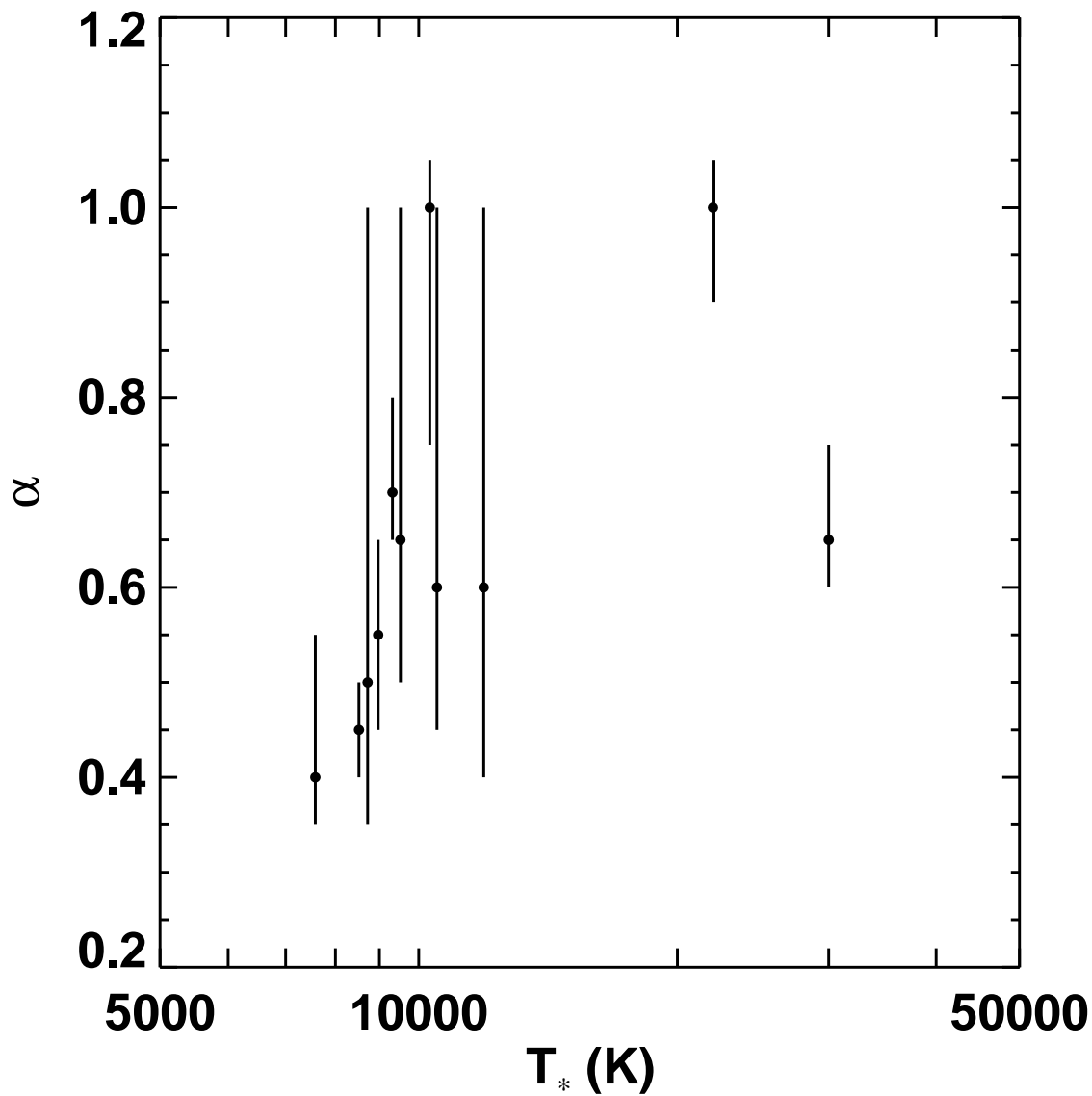}
\caption{Best-fit values of $\alpha$ for 
disk models with a power-law temperature
gradient, as a function of stellar effective
temperature, $T_{\ast}$.  Although the uncertainties on $\alpha$ are large 
for some objects, the data suggest that disks around cooler stars
tend to have shallower temperature gradients than disks around hotter stars.
$T_{\ast}$ is based on the assigned spectral types from \citet{EISNER+04}.
For objects with the same $T_{\ast}$, we have introduced an arbitrary offset
of 200 K for ease of viewing.
\label{fig:alphas}}
\end{figure}

\begin{figure}
\plotone{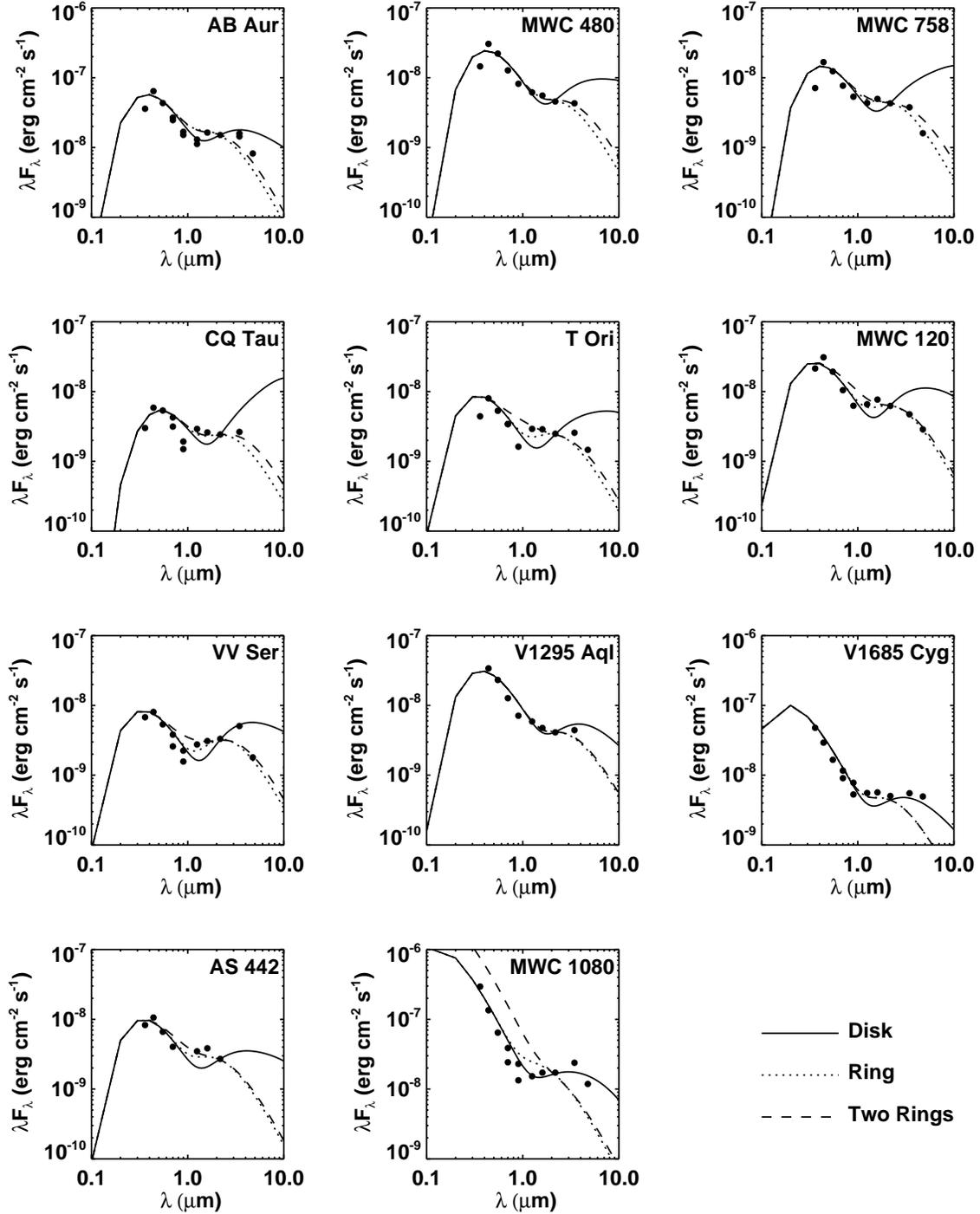}
\caption{SEDs computed for our best-fit models.
SED data from the literature \citep[][and references therein]{EISNER+04} are
indicated with filled circles, disk models with power-law temperature profiles
are shown with solid lines, single-temperature ring models are represented
by dotted lines, and two-ring models are shown as dashed lines.
\label{fig:seds}}
\end{figure}

We discern a striking correlation between temperature gradient and
effective temperature of the host star $T_{\ast}$, which may reflect 
relative differences in the dust/gas components or variations in the gradients
within individual components.  Using our power-law
disk model,
$T=T_{\rm in} (R/R_{\rm in})^{-\alpha}$, to illustrate this trend, 
we plot $\alpha$ as a function of 
$T_{\ast}$ in Figure \ref{fig:alphas}.  Larger values of $\alpha$ (steeper
gradients) are associated with objects with earlier spectral types.
For MWC 480, MWC 758, CQ Tau, and T Ori, the objects in our 
sample with the latest spectral types, the best-fit models require $\alpha =$
0.40--0.55.  In contrast, inferred values of $\alpha$ 
for other objects in our sample range between 0.60 and 1.0.

There is some degeneracy between temperature profiles
and optical depth profiles.  In our modeling we assumed the emission
was optically thick.  The fitted inner disk and ring temperatures may be 
underestimated to the extent that the optical depth of the emitting region,
$\tau_{\nu}$, is $< 1$. By the same token,
radial gradients in $\tau_{\nu}$ are somewhat 
degenerate with temperature gradients.
However, since $\tau_{\nu} $ is likely to decrease 
with increasing radius and 
wavelength (e.g., $\tau_{\nu} \propto R^{-1}$ for the surface layers of
geometrically thin disks; Chiang \& Goldreich 1997), 
$\tau_{\nu}$ gradients are more likely to produce larger sizes at 
shorter wavelengths (since more material at larger radii 
becomes optically thick). The gradients inferred from our data are therefore
likely tracing temperature profiles as opposed to optical depth profiles.

%See \citet{VDD05} for comparison of CG and DDN model predictions for the
%disk intensity profile.  It looks like DDN models produce steeper profiles
%than CG models.  Not sure this makes sense.

%at the small radii probed by our interferometric observations. These
%small values of $\alpha$ may indicate that we are actually observing 
%temperature gradients across the puffed-up inner disk edges.  In fact, 
%measurements of size versus wavelength may provide the only way to truly
%measure such temperature gradients.

\section{CONCLUSIONS}
We presented near-IR interferometry data, spectrally dispersed across 
5 channels from 2.0 to 2.4 $\mu$m,
from the Palomar Testbed Interferometer.  These data were used to constrain 
the temperature profiles of sub-AU-sized regions of 11 Herbig 
Ae/Be disks.  We found that a single-temperature ring of emission
did not fit the spectrally dispersed data well.  Two rings at different
stellocentric radii and with different temperatures provide better fits 
to the data.  The best fits (i.e., yielding the lowest reduced $\chi^2$ values)
are for disk models incorporating smooth radial temperature gradients.  
%We determined the inner disk 
%truncation radius, inner disk temperature, and the power-law
%exponent of the radial temperature profile for each disk model.

%The inner radii and temperatures are
%generally compatible with the dust disks being truncated by sublimation.
However, the smooth gradient 
disk models do not match broadband SEDs nearly as well as
two-ring models.  We therefore speculate that multiple emission components,
probably dust and gas, at different temperatures and stellocentric radii
are contributing to the
observed data.  
%The fact that our disk models provide the lowest reduced
%$\chi^2$ values for the 2.0--2.4 $\mu$m $V^2$ and flux data suggests that
More complex two-component models, including temperature gradients in one
or both components, will better match the observations.

The inferred temperature profiles are systematically shallower for disks
around less massive stars.  However, it is unclear whether this indicates
relative differences in the properties of the 
two components (gas and dust) as a function
of spectral type, or whether this trend traces temperature gradients
within a single component.  
%Around stars with spectral types later than A1,
%the temperature for our blackbody disk models declines with radius with
%an exponent of $\alpha \sim 0.5$, compatible with the exponent computed for
%recent models of puffed-up inner disk rims \citep{IN05}.  In contrast, disks 
%around earlier spectral type objects have $\alpha \ga 0.6$, which may be more 
%consistent with geometrically thin disk models.  
%Although these $\alpha$ values 
%could change if the radial dependence of optical depth is included in the
%models, the difference between early and late spectral type sources would
%remain.  We speculate that this difference is caused by higher inner disk
%gas masses, possibly associated with higher accretion rates, 
%in more massive sources.  These large gas masses may block stellar 
%optical/UV photons 
%needed to heat, and thereby puff up, the inner disk edges around high-mass
%HAEBE stars.

We also investigated whether gaseous emission, potentially at different 
spatial scales than the dust emission, contributes spectral line flux 
to our measurements.
Because CO and H$_2$O are expected to emit primarily in the edge channels
of our spectrally dispersed measurements, we performed fits of our disk models
to only the central channels.
Some objects showed a slight reduction (and in the case of AB Aur,
a substantial reduction) in the reduced $\chi^2$ values of the best-fit models
when only the central channels were used.  However, the best-fit parameters
for fits to the central channels and those for
fits to all data are consistent within
the 1$\sigma$ uncertainties for all objects.  
We are therefore unable to claim strong evidence for spectral line emission in 
the inner disks of our sample based on our current data.  Given the evidence
for hot inner disk continuum emission, future observations with better spectral
resolution and sensitivity are warranted to further investigate inner disk 
gas in these objects.

\noindent{\bf Acknowledgments.} The near-IR interferometry
data presented in this paper were obtained at the Palomar Observatory using 
the Palomar Testbed Interferometer, which is supported by NASA contracts to 
the Jet Propulsion Laboratory.  Science operations with PTI are possible 
through the efforts of the PTI Collaboration 
({\tt http://pti.jpl.nasa.gov/ptimembers.html}) and Kevin Rykoski. 
This research made use of software from 
the Michelson Science Center at the California Institute of Technology,
and data products from 2MASS.
J.A.E. is supported by a Miller Research Fellowship, E.I.C. 
acknowledges support from an Alfred P. Sloan Fellowship, and
B.F.L. acknowledges support from a Pappalardo
Fellowship in Physics.

\clearpage
\begin{deluxetable}{lccccc}
%\tabletypesize{\small}
\tablewidth{0pt}
\tablecaption{Circumstellar-to-Stellar Flux Ratios from $2.0-2.4$ $\mu$m
\label{tab:fratios}}
\tablehead{\colhead{Source} & \colhead{$r_{\rm 2.0 \: \mu m}$}
 & \colhead{$r_{\rm 2.1 \: \mu m}$}  & \colhead{$r_{\rm 2.2 \: \mu m}$}
 & \colhead{$r_{\rm 2.3 \: \mu m}$}  & \colhead{$r_{\rm 2.4 \: \mu m}$}}
\startdata
AB Aur &   3.8 &   4.7 &   5.8 &   7.0 &   8.2 \\
MWC 480 &   2.0 &   2.6 &   3.2 &   3.9 &   4.6 \\
MWC 758 &   3.3 &   4.1 &   5.0 &   6.0 &   7.0 \\
CQ Tau &   2.5 &   3.3 &   4.2 &   5.1 &   6.1 \\
T Ori &   5.3 &   6.7 &   8.5 &  10.3 &  12.0 \\
MWC 120 &   4.9 &   6.3 &   7.8 &   9.5 &  11.2 \\
VV Ser &   6.5 &   8.2 &  10.6 &  13.1 &  15.6 \\
V1295 Aql &   1.5 &   2.0 &   2.6 &   3.3 &   4.0 \\
V1685 Cyg &   3.7 &   4.7 &   6.2 &   7.6 &   9.1 \\
AS 442 &   3.6 &   4.7 &   6.2 &   7.7 &   9.1 \\
MWC 1080 &   5.2 &   7.0 &   9.1 &  11.6 &  14.1 \\
\enddata
\tablecomments{$r_{\lambda}$ denotes the ratio of the circumstellar
excess emission to the stellar photospheric emission. This ratio is
calculated for each of the PTI spectral channels, and typical uncertainties
are $\sim 20$--30\% (\S \ref{sec:ratios}).} 
\end{deluxetable}

\begin{deluxetable}{lcccc}
%\tabletypesize{\small}
\tablewidth{0pt}
\tablecaption{Best fits for single-temperature ring models
\label{tab:rings}}
\tablehead{\colhead{Source} & \colhead{$\chi_{\rm r}^2$} & 
\colhead{$\theta_{\rm in}/2$ (mas)}  & \colhead{$R_{\rm in}$ (AU)} & 
\colhead{$T_{\rm ring}$ (K)}}
\startdata
AB Aur & $2.346$ & $1.64_{-0.01}^{+0.01}$ & $0.23_{-0.01}^{+0.01}$ & $ 1860.0_{-  60.0}^{+  60.0}$ \\
MWC 480 & $1.153$ & $1.51_{-0.01}^{+0.01}$ & $0.21_{-0.01}^{+0.01}$ & $ 1440.0_{-  40.0}^{+  40.0}$ \\
MWC 758 & $2.267$ & $1.30_{-0.01}^{+0.01}$ & $0.20_{-0.01}^{+0.01}$ & $ 1600.0_{-  50.0}^{+  40.0}$ \\
CQ Tau & $1.179$ & $1.36_{-0.01}^{+0.02}$ & $0.20_{-0.01}^{+0.01}$ & $ 1420.0_{-  40.0}^{+  30.0}$ \\
T Ori & $0.904$ & $0.87_{-0.03}^{+0.02}$ & $0.39_{-0.01}^{+0.01}$ & $ 1640.0_{-  60.0}^{+  50.0}$ \\
MWC 120 & $1.173$ & $1.52_{-0.01}^{+0.01}$ & $0.76_{-0.01}^{+0.01}$ & $ 1560.0_{-  50.0}^{+  40.0}$ \\
VV Ser & $1.164$ & $1.44_{-0.03}^{+0.02}$ & $0.45_{-0.01}^{+0.01}$ & $ 1500.0_{-  50.0}^{+  40.0}$ \\
V1295 Aql & $1.005$ & $1.70_{-0.01}^{+0.01}$ & $0.49_{-0.01}^{+0.01}$ & $ 1320.0_{-  40.0}^{+  20.0}$ \\
V1685 Cyg & $4.101$ & $1.16_{-0.01}^{+0.01}$ & $1.16_{-0.01}^{+0.01}$ & $ 1760.0_{-  60.0}^{+  50.0}$ \\
AS 442 & $1.125$ & $0.87_{-0.04}^{+0.03}$ & $0.52_{-0.02}^{+0.02}$ & $ 1840.0_{-  80.0}^{+  80.0}$ \\
MWC 1080 & $1.840$ & $1.34_{-0.01}^{+0.01}$ & $1.34_{-0.01}^{+0.01}$ & $ 2300.0_{- 110.0}^{+  90.0}$ \\
\enddata
\tablecomments{ {Linear} inner ring radii, $R_{\rm in}$, are computed
using the derived angular inner radii, $\theta_{\rm in}/2$, and the distances
employed by \citet{EISNER+04}.}
\end{deluxetable}

\begin{deluxetable}{lccccc}
%\tabletypesize{\small}
\tablewidth{0pt}
\tablecaption{Best fits for two-ring models
\label{tab:tworings}}
\tablehead{\colhead{Source} & \colhead{$\chi_{\rm r}^2$} & 
\colhead{$T_{\rm ring,1}$ (K)} & \colhead{$\theta_{\rm in,2}/2$ (mas)}  & 
\colhead{$R_{\rm in,2}$ (AU)} & \colhead{$T_{\rm ring,2}$ (K)}}
\startdata
AB Aur & $1.769$ & $ 2310.0_{- 150.0}^{+ 140.0}$ & $2.14_{-0.09}^{+0.09}$ & $0.30_{-0.01}^{+0.01}$ & $ 1460.0_{-  70.0}^{+  60.0}$ \\
MWC 480 & $0.841$ & $ 1710.0_{-  90.0}^{+  80.0}$ & $2.13_{-0.16}^{+0.11}$ & $0.30_{-0.02}^{+0.02}$ & $ 1130.0_{-  50.0}^{+  60.0}$ \\
MWC 758 & $1.256$ & $ 2010.0_{-  90.0}^{+  90.0}$ & $2.24_{-0.13}^{+0.01}$ & $0.34_{-0.02}^{+0.01}$ & $ 1060.0_{-  20.0}^{+  40.0}$ \\
CQ Tau & $1.107$ & $ 1760.0_{- 140.0}^{+  80.0}$ & $2.20_{-0.41}^{+0.04}$ & $0.33_{-0.06}^{+0.01}$ & $ 1010.0_{-  10.0}^{+ 130.0}$ \\
T Ori & $1.139$ & $ 3100.0_{- 920.0}^{+ 690.0}$ & $1.24_{-0.24}^{+0.53}$ & $0.56_{-0.11}^{+0.24}$ & $ 1220.0_{- 220.0}^{+ 270.0}$ \\
MWC 120 & $0.868$ & $ 3850.0_{- 880.0}^{+ 140.0}$ & $1.83_{-0.16}^{+0.11}$ & $0.92_{-0.08}^{+0.06}$ & $ 1350.0_{-  80.0}^{+ 100.0}$ \\
VV Ser & $1.111$ & $ 2740.0_{- 460.0}^{+ 310.0}$ & $1.91_{-0.24}^{+0.23}$ & $0.59_{-0.07}^{+0.07}$ & $ 1200.0_{-  90.0}^{+ 130.0}$ \\
V1295 Aql & $1.174$ & $ 1500.0_{-  10.0}^{+ 320.0}$ & $1.78_{-0.02}^{+0.12}$ & $0.52_{-0.01}^{+0.03}$ & $ 1270.0_{-  60.0}^{+  40.0}$ \\
V1685 Cyg & $4.195$ & $ 1710.0_{- 210.0}^{+2280.0}$ & $1.16_{-0.01}^{+0.05}$ & $1.17_{-0.01}^{+0.05}$ & $ 1750.0_{-  90.0}^{+  80.0}$ \\
AS 442 & $1.187$ & $ 3580.0_{- 640.0}^{+ 410.0}$ & $1.00_{-0.01}^{+0.12}$ & $0.60_{-0.01}^{+0.07}$ & $ 1590.0_{- 180.0}^{+  70.0}$ \\
MWC 1080 & $1.224$ & $27060.0_{-8000.0}^{+5540.0}$ & $1.73_{-0.12}^{+0.07}$ & $1.73_{-0.13}^{+0.07}$ & $ 1740.0_{- 250.0}^{+  90.0}$ \\
\enddata
\tablecomments{ 
Values of $R_{\rm in,2}$, are computed
using the derived angular inner radii, $\theta_{\rm in,2}/2$, and the distances
employed by \citet{EISNER+04}.  
We have assumed an inner ring radius $R_{\rm in,1}$=0.1 AU for these models;
the high inner ring temperature for MWC 1080 may indicate that a larger value 
of $R_{\rm in,1}$ is warranted for this source (\S \ref{sec:mod2rings}).}
\end{deluxetable}

\begin{deluxetable}{lcccccc}
\tabletypesize{\small}
\tablewidth{0pt}
\tablecaption{Best fits for disk models with temperature gradients
\label{tab:results}}
\tablehead{\colhead{Source} & \colhead{$T_{\ast}$ (K)} &
\colhead{$\chi_{\rm r}^2$} & 
\colhead{$\theta_{\rm in}/2$ (mas)} & \colhead{$R_{\rm in}$ (AU)} &
\colhead{$\alpha$} & \colhead{$T_{\rm in}$ (K)}}
\startdata
\multicolumn{7}{c}{Best-Fit Disk Models} \\
\hline
CQ Tau & 7580 & $0.976$ & $0.60_{-0.18}^{+0.29}$ & $0.09_{-0.03}^{+0.04}$ & $ 0.40_{-0.05}^{+0.15}$ & $ 1360.0_{-  60.0}^{+  80.0}$ \\
MWC 758 & 8720 & $1.117$ & $0.63_{-0.12}^{+0.02}$ & $0.09_{-0.02}^{+0.01}$ & $ 0.45_{-0.05}^{+0.05}$ & $ 1490.0_{-  50.0}^{+  50.0}$ \\
T Ori & 8720 & $0.759$ & $0.42_{-0.32}^{+0.34}$ & $0.19_{-0.14}^{+0.15}$ & $ 0.50_{-0.15}^{+0.50}$ & $ 1580.0_{-  80.0}^{+ 550.0}$ \\
MWC 480 & 8970 & $0.780$ & $1.03_{-0.15}^{+0.11}$ & $0.14_{-0.02}^{+0.01}$ & $ 0.55_{-0.10}^{+0.10}$ & $ 1340.0_{-  50.0}^{+  40.0}$ \\
AB Aur & 9520 & $1.657$ & $1.17_{-0.06}^{+0.09}$ & $0.16_{-0.01}^{+0.01}$ & $ 0.70_{-0.05}^{+0.10}$ & $ 1700.0_{-  70.0}^{+  80.0}$ \\
VV Ser & 9520 & $1.029$ & $1.00_{-0.22}^{+0.24}$ & $0.31_{-0.07}^{+0.07}$ & $ 0.65_{-0.15}^{+0.35}$ & $ 1430.0_{-  50.0}^{+  80.0}$ \\
MWC 120 & 10500 & $0.524$ & $1.06_{-0.22}^{+0.26}$ & $0.53_{-0.11}^{+0.13}$ & $ 0.60_{-0.15}^{+0.40}$ & $ 1440.0_{-  70.0}^{+ 120.0}$ \\
V1295 Aql & 10500 & $1.068$ & $1.51_{-0.12}^{+0.02}$ & $0.44_{-0.04}^{+0.01}$ & $ 1.00_{-0.25}^{+0.05}$ & $ 1330.0_{-  70.0}^{+  30.0}$ \\
AS 442 & 11900 & $1.107$ & $0.49_{-0.38}^{+0.26}$ & $0.29_{-0.23}^{+0.16}$ & $ 0.60_{-0.20}^{+0.40}$ & $ 1750.0_{-  90.0}^{+ 730.0}$ \\
V1685 Cyg & 22000 & $4.158$ & $0.96_{-0.04}^{+0.01}$ & $0.97_{-0.05}^{+0.01}$ & $ 1.00_{-0.10}^{+0.05}$ & $ 1710.0_{-  60.0}^{+  70.0}$ \\
MWC 1080 & 30000 & $1.119$ & $0.79_{-0.08}^{+0.10}$ & $0.79_{-0.08}^{+0.11}$ & $ 0.65_{-0.05}^{+0.10}$ & $ 2060.0_{- 100.0}^{+  90.0}$ \\
\hline
\multicolumn{7}{c}{``Continuum''--Only Fits} \\
\hline
CQ Tau & 7580 & $1.078$ & $1.16_{-0.46}^{+0.04}$ & $0.17_{-0.07}^{+0.01}$ & $ 1.00_{-0.55}^{+0.05}$ & $ 1440.0_{- 140.0}^{+  60.0}$ \\
MWC 758 & 8720 & $1.218$ & $0.71_{-0.19}^{+0.18}$ & $0.11_{-0.03}^{+0.03}$ & $ 0.50_{-0.10}^{+0.15}$ & $ 1500.0_{-  90.0}^{+  70.0}$ \\
T Ori & 8720 & $0.648$ & $0.23_{-0.13}^{+0.51}$ & $0.11_{-0.06}^{+0.23}$ & $ 0.40_{-0.05}^{+0.60}$ & $ 1740.0_{- 270.0}^{+ 410.0}$ \\
MWC 480 & 8970 & $0.972$ & $0.96_{-0.19}^{+0.24}$ & $0.14_{-0.03}^{+0.03}$ & $ 0.50_{-0.10}^{+0.25}$ & $ 1330.0_{-  70.0}^{+  80.0}$ \\
AB Aur & 9520 & $0.712$ & $1.16_{-0.23}^{+0.19}$ & $0.16_{-0.03}^{+0.03}$ & $ 0.70_{-0.20}^{+0.30}$ & $ 1700.0_{- 140.0}^{+ 190.0}$ \\
VV Ser & 9520 & $0.709$ & $1.03_{-0.27}^{+0.21}$ & $0.32_{-0.09}^{+0.06}$ & $ 0.70_{-0.20}^{+0.30}$ & $ 1450.0_{- 100.0}^{+ 130.0}$ \\
MWC 120 & 10500 & $0.737$ & $1.26_{-0.54}^{+0.04}$ & $0.63_{-0.27}^{+0.02}$ & $ 1.00_{-0.60}^{+0.05}$ & $ 1570.0_{- 210.0}^{+  70.0}$ \\
V1295 Aql & 10500 & $1.126$ & $1.54_{-0.33}^{+0.03}$ & $0.45_{-0.10}^{+0.01}$ & $ 1.00_{-0.50}^{+0.05}$ & $ 1320.0_{- 130.0}^{+  60.0}$ \\
AS 442 & 11900 & $0.978$ & $0.10_{-0.00}^{+0.34}$ & $0.06_{-0.00}^{+0.20}$ & $ 0.35_{-0.05}^{+0.15}$ & $ 2260.0_{- 590.0}^{+  80.0}$ \\
V1685 Cyg & 22000 & $5.582$ & $1.00_{-0.20}^{+0.01}$ & $1.00_{-0.21}^{+0.02}$ & $ 1.00_{-0.35}^{+0.05}$ & $ 1690.0_{- 120.0}^{+ 100.0}$ \\
MWC 1080 & 30000 & $1.008$ & $0.83_{-0.20}^{+0.20}$ & $0.84_{-0.20}^{+0.21}$ & $ 0.70_{-0.15}^{+0.30}$ & $ 2060.0_{- 150.0}^{+ 220.0}$ \\
\enddata
\tablecomments{The stellar effective temperatures, $T_{\ast}$, are based on
the assigned spectral types from \citet{EISNER+04}.  Linear inner radii,
$R_{\rm in}$, are computed
using the derived angular inner radii, $\theta_{\rm in}/2$, and the distances
employed by \citet{EISNER+04}.  Objects are listed in order of increasing
$T_{\ast}$ to illustrate the correlation of $T_{\ast}$ with $R_{\rm in}$,
$\alpha$, and $T_{\rm in}$.}
\end{deluxetable}

\clearpage

\epsscale{0.8}
\clearpage
\epsscale{0.9}
%\begin{figure}
{\plotone{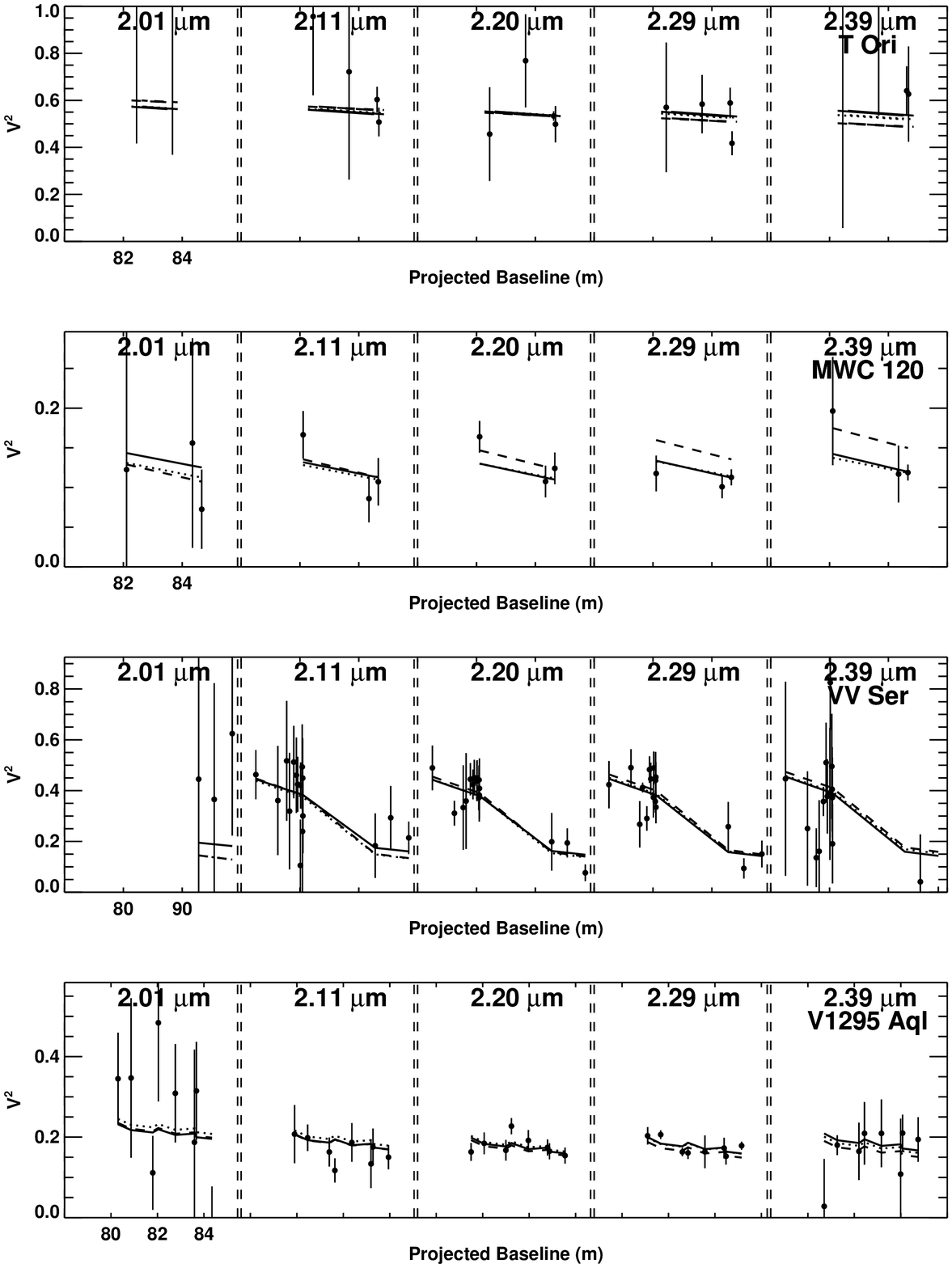}}\\[5mm]
\centerline{Fig. 4. --- continued.}
%\caption{Figure \ref{fig:modfits} continued.
%\label{fig:modfits2}}
%\end{figure}
\clearpage
%\begin{figure}
{\plotone{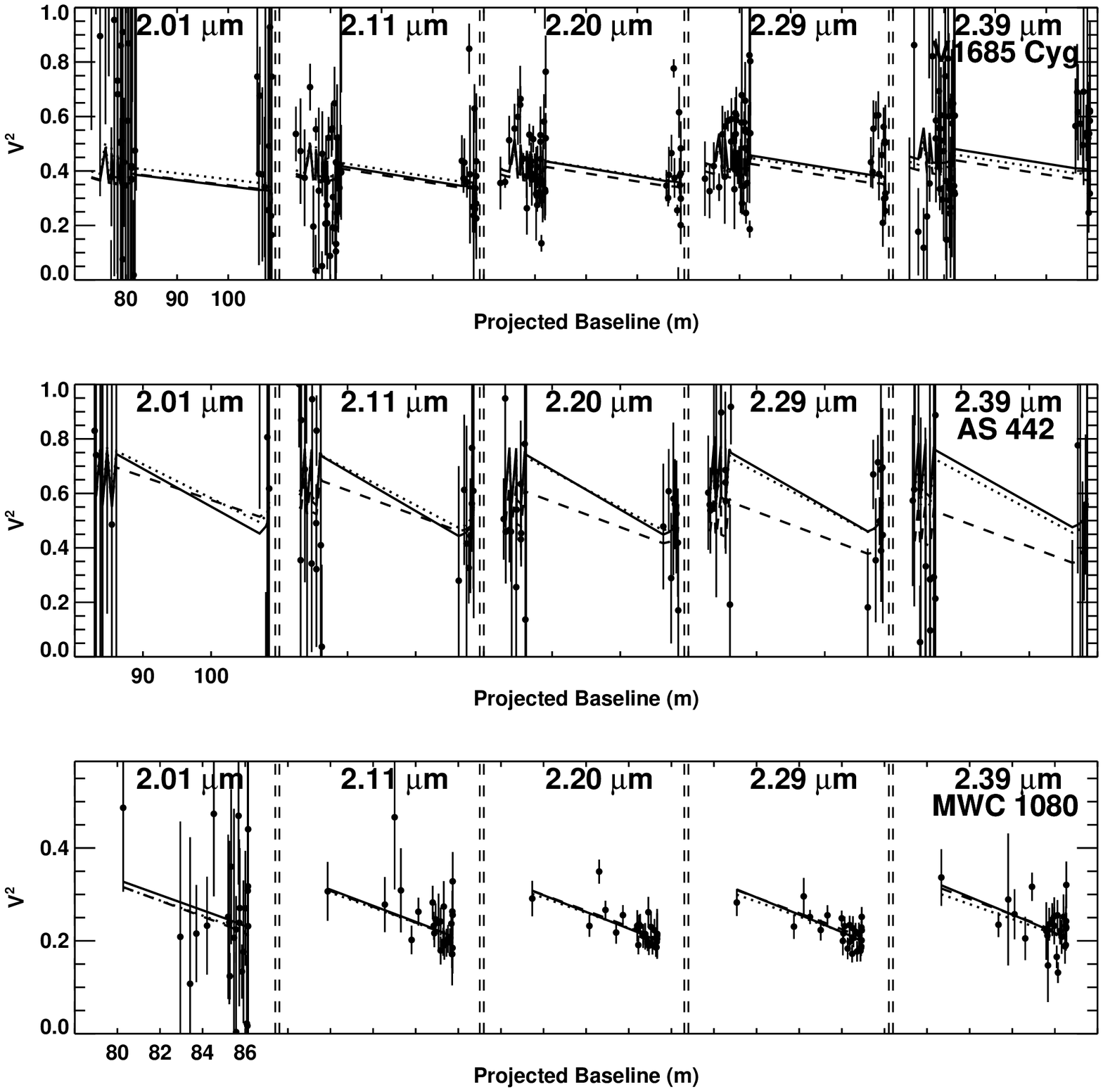}}\\[5mm]
\centerline{Fig. 4. --- continued.}
%\caption{Figure \ref{fig:modfits} continued.
%\label{fig:modfits3}}
%\end{figure}
\clearpage

%\epsscale{0.8}
%\begin{figure}
%\plotone{figs/specmodels.epsi}
%\caption{$V^2$ as a function of wavelength  (points) 
%and simple models (line).  Each $V^2$ point is the weighted mean of all 
%measurements with a single baseline (NW) for a given spectral channel, and 
%the error bar is given by the weighted standard deviation of the data.  
%\label{fig:modexample}}
%\end{figure}

\end{document}